\documentclass[]{aa}  

\usepackage{xcolor}
\usepackage{graphicx}
\usepackage{txfonts}
\usepackage{subfig}        

\usepackage{lscape}             
                               
\usepackage{placeins}           
                               
\usepackage{natbib,twoopt}
\usepackage[breaklinks=true]{hyperref} 
\bibpunct{(}{)}{;}{a}{}{,}             
	
\makeatletter
  \newcommandtwoopt{\citeads}[3][][]{\href{http://adsabs.harvard.edu/abs/#3}%
    {\def\hyper@linkstart##1##2{}%
     \let\hyper@linkend\@empty\citealp[#1][#2]{#3}}}
  \newcommandtwoopt{\citepads}[3][][]{\href{http://adsabs.harvard.edu/abs/#3}%
    {\def\hyper@linkstart##1##2{}%
     \let\hyper@linkend\@empty\citep[#1][#2]{#3}}}
  \newcommandtwoopt{\citetads}[3][][]{\href{http://adsabs.harvard.edu/abs/#3}%
    {\def\hyper@linkstart##1##2{}%
     \let\hyper@linkend\@empty\citet[#1][#2]{#3}}}
  \newcommandtwoopt{\citeyearads}[3][][]%
    {\href{http://adsabs.harvard.edu/abs/#3}
    {\def\hyper@linkstart##1##2{}%
     \let\hyper@linkend\@empty\citeyear[#1][#2]{#3}}}
\makeatother

\begin{document}
   \title{Baryonic assembly bias in X-ray-selected galaxy groups and clusters: Insights from the Magneticum simulation}
   \titlerunning{Baryonic assembly bias in X-ray-selected galaxy groups and clusters}

   \author{Ilaria~Marini
          \inst{1, 2, 3}\thanks{Corresponding author: \texttt{ilaria.marini@eso.org}}
          \and
          Tiago~Castro\inst{4, 5, 6, 7, 8}
          \and
          Paola~Popesso\inst{1, 2}
          \and
          Klaus~Dolag\inst{3, 9, 2}
          \and
          Veronica~Biffi\inst{5, 6}
          \and\\
          Natanael~de~Isídio\inst{1}
          \and 
          Daudi~Mazengo\inst{1, 10}
          \and
          Victoria~Toptun\inst{1}
          }
   \institute{European Southern Observatory, Karl-Schwarzschild-Straße 2, 85748, Garching bei München, Germany
         \and
            Excellence Cluster ORIGINS, Boltzmannstr. 2, D-85748 Garching bei M\"unchen, Germany
        \and 
            Universitäts-Sternwarte, Fakultät für Physik, Ludwig-Maximilians-Universität München, Scheinerstr.1, 81679 München, Germany
        \and
            Department of Mathematical Physics, Institute of Physics, University of S\~ao Paulo, R. do Mat\~ao 1371, 05508-090, S\~ao Paulo, SP, Brazil
        \and
            INAF – Osservatorio Astronomico di Trieste, Via Tiepolo 11, 34143 Trieste, Italy
        \and 
            IFPU – Institute for Fundamental Physics of the Universe, Via Beirut 2, I-34014 Trieste, Italy
        \and
            INFN, Sezione di Trieste, Via Valerio 2, 34127 Trieste TS, Italy        
        \and
            ICSC - Centro Nazionale di Ricerca in High Performance Computing, Big Data e Quantum Computing, Via Magnanelli 2, Bologna, Italy        
        \and 
            Max-Planck-Institut für Astrophysik, Karl-Schwarzschildstr. 1, 85741 Garching bei M\"unchen, Germany
        \and
             Department of Physics, College of Natural and Mathematical Sciences, University of Dodoma, P.O. Box 338, Dodoma, Tanzania
             }
   \date{Received 23 March 2026 ; accepted 03 July 2026.}

  \abstract
   {Galaxy groups and clusters trace the large-scale matter distribution, with their clustering usually interpreted mainly as a function of halo mass. Yet, at fixed mass, their baryonic properties retain information about halo growth, gas accretion, and feedback. The intrinsic scatter in X-ray luminosity and gas fraction suggests that X-ray-selected systems may not be a random subset of the halo population. If these observables correlate with halo assembly, they may trace secondary variations in halo bias. We test this using the Magneticum simulation, measuring the clustering of systems selected by X-ray luminosity and gas fraction at fixed halo mass. 

    We constructed mass-matched halo subsamples by ranking systems in percentiles according to their X-ray luminosity, and we derived the linear halo-matter bias. We found that halos with higher X-ray luminosity are more strongly clustered than those with lower luminosity. For the fiducial 84th--16th percentile split, the large-scale bias difference is $\Delta b_{\rm lin}=0.17\pm0.03$, corresponding to a $\sim17\%$ enhancement relative to the X-ray-faint sample. A less extreme 67th--33rd percentile split gives a consistent signal, with $\Delta b_{\rm lin}=0.12\pm0.02$, corresponding to a $\sim12\%$ clustering enhancement. When analysed as a function of halo mass, differences in the large-scale bias are of the order of 10–30\% between the two samples. The signal is strongest at galaxy group scales ($10^{13}-10^{14} M_{\odot}$) and negligible for cluster-size halos ($>10^{14} M_{\odot}$). The complementary analysis, based on the gas fraction, reveals an even stronger clustering dependence on baryonic content: a $\sim39\%$ ($\sim 26\%)$ relative clustering enhancement in the 84th-16th (67th-33rd) sample.  Baryonic assembly bias is already present at high redshift (from $z\simeq2$), but X-ray luminosity becomes a significant tracer of it ($>5 \sigma$) only at lower redshift ($z\simeq0.3$), once the thermodynamic state of the gas is more closely coupled to baryon retention. Matching halos simultaneously by mass and formation time reduces the large-scale bias difference to below $2\sigma$. This indicates that formation time captures a substantial part of the luminosity- and gas-fraction-dependent clustering signals in Magneticum.
    
    These results show that in the Magneticum simulation suite X-ray luminosity traces a secondary dependence of halo clustering beyond mass, providing a baryonic manifestation of the halo assembly bias. }


   \maketitle

\section{Introduction}
The thermodynamic properties of the intracluster medium and intragroup medium (ICM and IGM) as well as the general baryonic properties of galaxy groups and clusters reflect the complex interplay between gravitational collapse and non-gravitational processes such as cooling and feedback \citep{borgani_effect_2002, borgani_thermodynamical_2008, voit_observationally_2005, voit_tracing_2005}. While the halo mass sets the overall scale of these systems \citep{kaiser_evolution_1986, bryan_statistical_1998}, observables such as the X-ray luminosity, integrated Sunyaev-Zeldovich (SZ) signal, $Y$, and the hot gas fraction show substantial scatter at fixed mass \citep{motl_integrated_2005, kravtsov_new_2006, sun_chandra_2009, pratt_galaxy_2009, mantz_cosmology_2016,  eckert_low-scatter_2020}. This serves as an indication that non-gravitational processes play a central role in shaping the IGM and ICM. 

A growing body of work suggests that these baryonic properties are closely linked to halo accretion history, merger activity, and environment. In the Magneticum simulation suite, X-ray-bright galaxy groups and clusters correspond to late-forming halos that retain large hot-gas reservoirs, while early-forming systems are more gas-depleted and X-ray-faint \citep{marini_impact_2025}. Other simulation suites, such as Hyenas and Flamingo, report the opposite trend, linking enhanced X-ray luminosity to early formation times and higher halo concentrations \citep{cui_hyenas_2024,costello_flamingo_2025}. The diversity of these predictions underscores that the physical origin of the scatter in the $L_X$--$M$ relation remains unsettled \citep{aljamal_mass_2025,braspenning_origin_2025}. \textcolor{black}{ At the same time, this disagreement is potentially informative: if different models predict different connections between X-ray properties and halo assembly, then X-ray observables may provide a way to discriminate between the galaxy-formation physics implemented in contemporary hydrodynamical simulations.}

If X-ray luminosity and gas fraction trace these secondary halo properties, then X-ray selection should preferentially select systems with distinct assembly histories and, consequently, distinct large-scale clustering \citep[e.g.][]{andreon_amazing_2016, andreon_why_2019}. At fixed mass, X-ray–bright and X-ray–faint systems may therefore occupy different environments and exhibit enhanced or suppressed large-scale bias relative to the mass-selected average \citep{popesso_x-ray_2024, zarattini_where_2024}.

This question naturally connects to the framework of halo assembly bias. In the standard theory of halo clustering, the large-scale bias is primarily a function of halo mass \citep{mo_analytic_1996,sheth_large-scale_1999}. Numerical studies have shown that this description is incomplete: halos of identical mass but different assembly histories can exhibit systematically different clustering amplitudes, a phenomenon known as halo assembly bias \citep{gao_age_2005,wechsler_dependence_2006,jing_dependence_2007}. Assembly bias therefore provides a theoretical framework for understanding how secondary halo properties, such as concentration, accretion rate, merger history, or gas content, become correlated with the large-scale environment \citep{galarraga-espinosa_properties_2021, manolopoulou_environmental_2021, gouin_gas_2022, zarattini_where_2024,Shreeram2026}.

This effect would have important implications for cosmological and astrophysical applications of X-ray–selected samples. Assembly bias in baryonic observables would affect the interpretation of clustering measurements, bias cosmological parameter constraints -- if left unaccounted for -- and require extensions to halo occupation and empirical galaxy–halo models \citep{allen_cosmological_2003, vikhlinin_chandra_2009,allen_cosmological_2011, mantz_weighing_2015, chiu_cosmological_2023, ghirardini_srgerosita_2024}. Unbiased scaling properties are key to revealing significant variations in the X-ray properties of groups, with some systems appearing bright and extended \citep{bulbul_erosita_2022, zheng_measuring_2023, li_robust_2024, popesso_x-ray_2024, popesso_average_2024, popesso_hot_2024}. Moreover, if the large-scale environment regulates gas inflow and retention, it may also influence the fuelling of central supermassive black holes, offering a way to discriminate between competing feedback and accretion prescriptions implemented in contemporary hydrodynamical simulations \citep{oppenheimer_simulating_2021}. 

\textcolor{black}{ In this work, we use the Magneticum simulations to predict the large-scale clustering of X-ray-selected halos and to investigate the physical origin of this signal.} This paper is organised as follows. Section~\ref{sec:methods} describes the data, sample selection, and clustering estimators. Section~\ref{sec:results} presents the clustering measurements of X-ray--bright and X-ray--faint systems at fixed mass, their dependence on halo mass and gas fraction, and their evolution with redshift. We then discuss the physical origin of the luminosity-dependent clustering signal and its connection to halo assembly and baryon retention in Sect.~\ref{sec:4.1}. Section~\ref{sec:conclusions} summarises our findings and discusses the implications for halo assembly bias, baryonic physics, and cosmological modelling.

\section{Methodology}
\label{sec:methods}
\subsection{The Magneticum simulations}
The Magneticum Pathfinder simulations \citep{dolag_encyclopedia_2025} are an extended set of the state-of-the-art cosmological hydrodynamical simulations carried out with P-GADGET3 \citep[an updated version of the public GADGET-2 code;][]{springel_cosmological_2003}. Key advancements include a higher order kernel function, time-dependent artificial viscosity, and artificial conduction schemes \citep{dolag_turbulent_2005,beck_improved_2016}.
\par
For this study, we used \textit{Box2/hr}, which follows the evolution of $2\times1584^{3}$ particles in a cosmological box of size $352 h {^{-1}}$ cMpc. The particle masses have the following resolutions: $m_\mathrm{DM} = 6.9\times10^8 h{^{-1}}$~M$_{\odot}$ and $m_\mathrm{gas} = 1.4\times10^8 h^{-1}$~M$_{\odot}$. The Plummer equivalent length for the DM particles corresponds to $\epsilon=3.75~h^{-1}$ ckpc, whereas gas, stars, and black hole particles retain $\epsilon=3.75~h{^{-1}}$~ckpc, $2~h {^{-1}}$~ckpc, and $2~h^{-1}$~ckpc at $z=0$, respectively. The simulation adopts the WMAP7 cosmology \citep{komatsu_hunting_2010}: $\Omega_\mathrm{M}=0.272$, $\Omega_\mathrm{bar}=0.046$, $n_s=0.963$, $\sigma_8 = 0.809$, and $H_0=100 \, h$ km~s${^{-1}}$~Mpc${^{-1}}$ with $h=0.704$. All logarithms are in base 10. \textcolor{black}{ We also performed a consistency check using the larger Magneticum \textit{Box2b/hr}, which has a side length of $640~h^{-1}$ cMpc. This larger volume provides a useful test of whether the gas-fraction-dependent clustering signal is affected by the finite volume of \textit{Box2/hr}. Owing to the limited number of available redshift snapshots in \textit{Box2b/hr}, which prevents us from calculating the halo formation time, and the fact that the full X-ray luminosity analysis has not yet been carried out for this box, we do not include these measurements in the present paper. Nevertheless, the gas-fraction split in \textit{Box2b/hr} yields the same qualitative behaviour as in \textit{Box2/hr}, with gas-rich systems showing a stronger large-scale clustering signal than gas-poor systems at fixed mass. This supports the interpretation that the baryonic assembly-bias signal reported here is not driven by the finite volume of the fiducial box.
}

\par
In brief, the simulations adopt a comprehensive suite of subgrid physics models to represent unresolved baryonic processes. These include metallicity-dependent radiative cooling \citep{wiersma_effect_2009}, a redshift-evolving UV background \citep{haardt_modelling_2001}, prescriptions for star formation and stellar-driven outflows \citep{springel_cosmological_2003}, and detailed chemical enrichment from stellar evolution, tracking the production of H, He, and a broad set of heavy elements \citep{tornatore_chemical_2007}. The growth and energetic feedback of supermassive black holes are modelled following established accretion and active galactic nucleus (AGN) feedback schemes \citep{springel_cosmological_2005, di_matteo_energy_2005, fabjan_simulating_2010, hirschmann_cosmological_2014}.
\par
Halos are identified using the \textsc{Subfind} algorithm \citep{springel_populating_2001, dolag_substructures_2009}. The procedure begins by locating parent halos through a standard friends-of-friends (FoF) algorithm with a linking length of $b = 0.16$ times the mean inter-particle separation. Then, the halo classification is refined via a gradient descent of the density field and an unbinding procedure, which ultimately identifies the halos and subhalos within them. The Magneticum simulations have been extensively validated against observational data. Previous works have demonstrated their ability to reproduce a broad range of galaxy properties and the thermodynamic and structural characteristics of galaxy clusters  \citep[see][for a review]{dolag_encyclopedia_2025}. Specifically, many studies have been devoted to the X-ray emission associated with galaxies and clusters \citep{biffi_investigating_2013, biffi_erosita_2022, veronica_erosita_2022, veronica_erosita_2024, seppi_offset_2023, scheck_hydrostatic_2023,  bogdan_circumgalactic_2023, zuhone_effects_2023, churazov_prospects_2023, vladutescu-zopp_decomposition_2023, vladutescu-zopp_radial_2025, Kruglov2025, toptun_erosita_2025, biffi_full_2025, toptun_shmr_2026}.

\subsection{The X-ray luminosity and gas properties}
X-ray luminosities are calculated on a particle basis. The X-ray emission is modelled using PHOX \citep{biffi_observing_2012, biffi_investigating_2013, biffi_agn_2018, vladutescu-zopp_decomposition_2023}, which simulates X-ray spectral emission for the gas, the black hole, and the star particles (i.e. X-ray binaries) using spectral models from the XSPEC library \cite[v12;][]{arnaud_xspec_1996}. For this work, we only used the X-ray luminosity coming from gas particles; thus, we focussed on the ICM only, centred on the central galaxy and integrated within a given aperture, which naturally includes emission from satellites. The hot gas emission is assumed to follow \texttt{vapec} \citep{smith_collisional_2001} with a single temperature model and in the presence of heavy elements, for which the single abundances are explicitly tracked. Solar abundances are assumed following \cite{anders_abundances_1989}. The parameters are derived from the chemical and thermal properties of the single gas particles in the simulations. A simple foreground absorber can be assumed with the \texttt{wabs} model \citep{morrison_interstellar_1983} with a column density of $N_{H}=10^{20}$ cm$^{-2}$. A fiducial collecting area, $A_{fid}$, and exposure time, $\tau_{fid}$, are initially assumed to generate a statistically significant number of photons from the spectra using a Monte Carlo approach. Photons are kept in the rest-frame $0.5-2.0$ keV energy range. This self-consistent description of the X-ray emission has granted a multitude of works on the topic \citep{vladutescu-zopp_decomposition_2023, biffi_erosita_2022, marini_detecting_2024, marini_impact_2025, marini_detecting_2025, toptun_erosita_2025, vladutescu-zopp_radial_2025, Groth2026}.

\begin{figure*}
    \centering
    \subfloat[]{
        \includegraphics[width=0.48\linewidth]{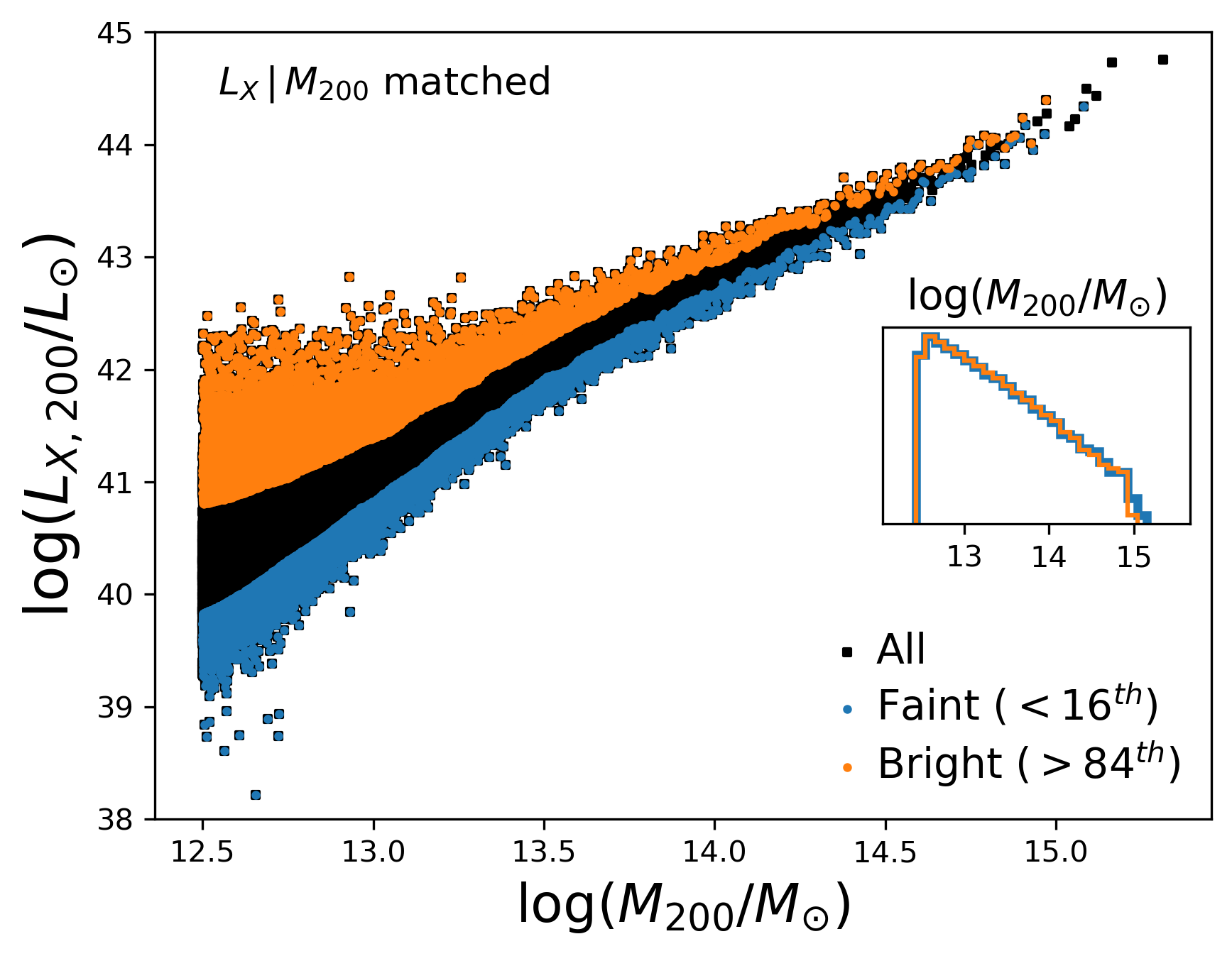}
    }
    \hfill
    \subfloat[]{
        \includegraphics[width=0.48\linewidth]{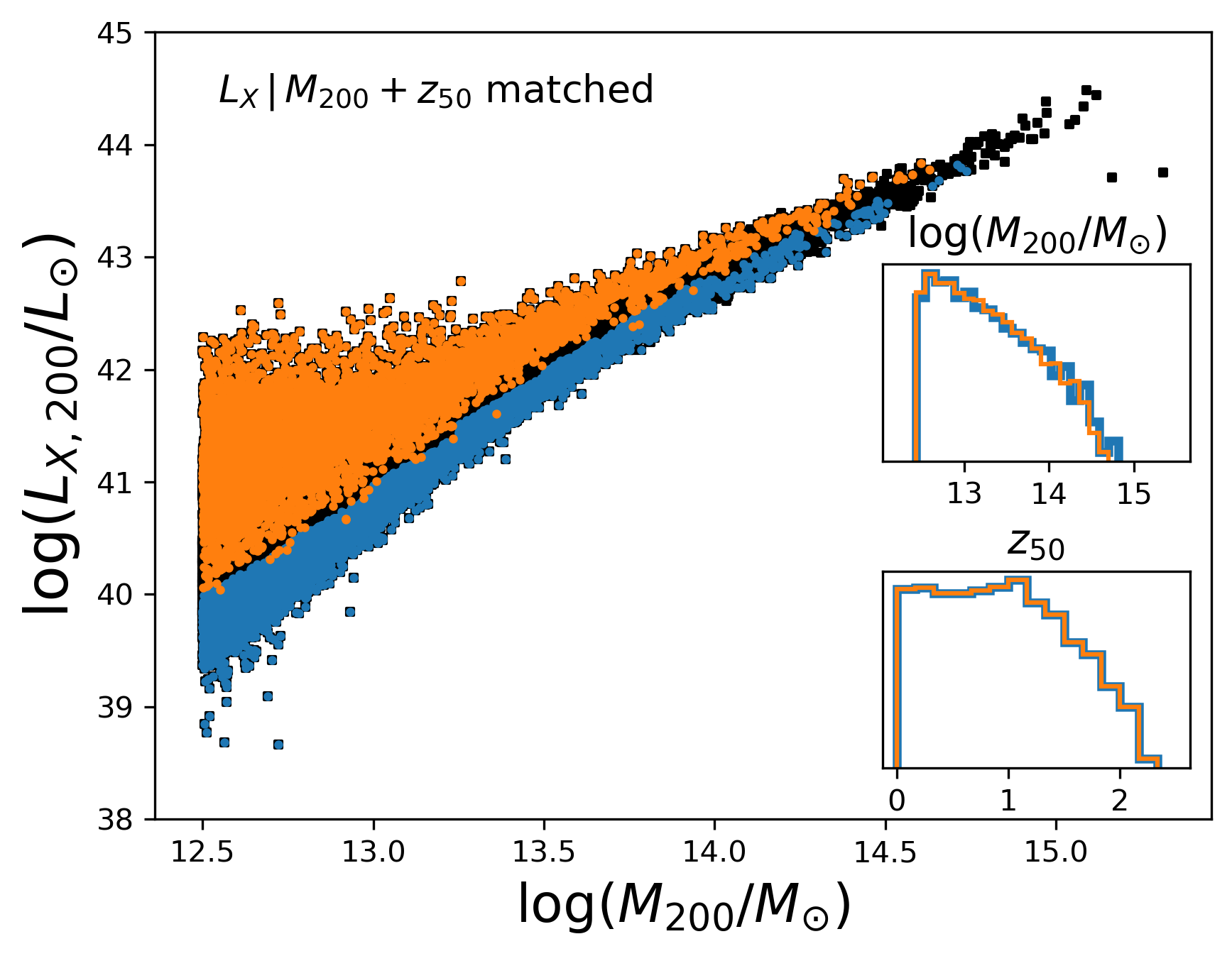}
    }
    \caption{Scaling relation between the X-ray luminosity within $R_{200}$ and the halo mass in Box2/hr at $z=0.03$ in the Magneticum simulation. Panel (a) Selection at fixed halo mass, with orange points indicating halos above the 84th percentile and blue points indicating halos below the 16th percentile. Panel (b) Corresponding selection after additionally matching in formation time. The inset histograms show the corresponding matched distributions: $M_{200}$ for the mass-matched selection, and both $M_{200}$ and $z_{50}$ for the formation-time-matched selection.}
    \label{fig:Lx_M}
\end{figure*}

\begin{figure}
    \centering    
    \subfloat{\includegraphics[width=\linewidth]{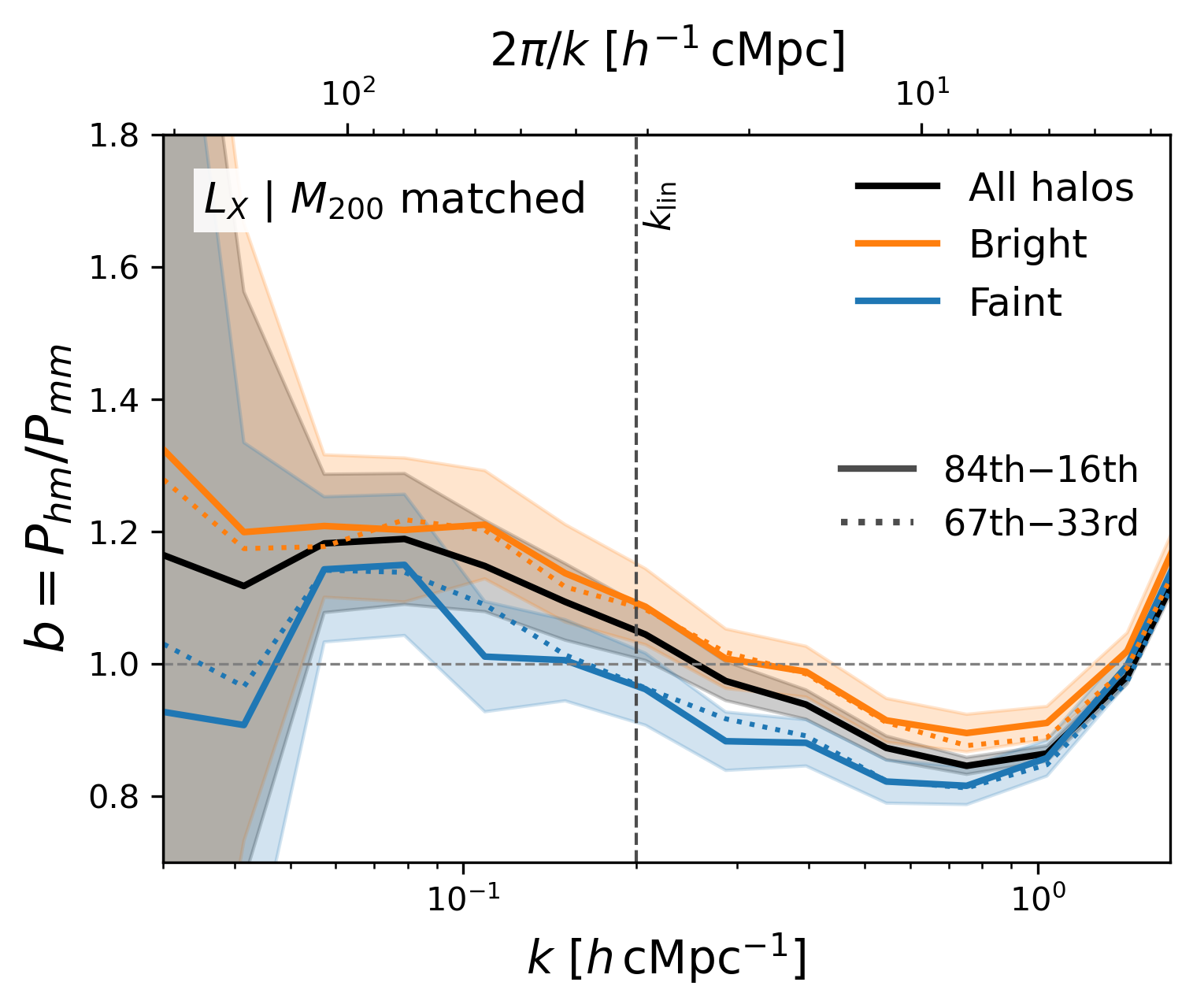}} 
    \caption{Scale-dependent bias for the full halo sample (black) as well as for X-ray-bright (orange) and X-ray-faint (blue) \textcolor{black}{ mass-matched} subsamples. Solid lines show the 84th--16th percentile split; dotted lines show the 67th--33rd percentile split. Shaded regions indicate the $1\sigma$ scatter from jackknife resampling. The vertical dashed line marks $k_{\rm lin}=0.2\,h\,{\rm cMpc}^{-1}$.}
    \label{fig:Pk}
\end{figure}
\subsection{Sample selection}
\label{sec:2.3}
We built our halo sample from the group catalogue of the Magneticum simulation box at the lowest redshift available (i.e. $z\simeq0.03$), which is complete above a minimum halo mass\footnote{We define $M_{\Delta}$ as the mass encompassed by a mean overdensity equal to $\Delta$ times the critical density of the universe $\rho_c(z)$.} of $M_{200,\min} = 10^{12.5}\,M_{\odot}$. From this parent sample, we identified X-ray-bright and X-ray-faint halos by ranking systems at fixed mass according to their position in the $L_{X,200}\!-\!M_{200}$ relation (see Fig.~\ref{fig:Lx_M}). Within narrow bins of $\log M_{200}$ (width 0.06 dex), we ranked halos by $L_{X, 200}$ and select halos above the 84th percentile and below the 16th percentile in each bin. We applied this procedure over the full mass range $10^{12.5} < M_{200}/M_{\odot} < 10^{15}$. The final bright and faint samples are the union over all mass bins, thereby isolating the tails of the luminosity scatter. As a robustness check, we repeat the analysis using the 67th and 33rd percentiles and find consistent clustering trends. 
\par
This percentile-based selection ensures that both subsamples share a similar halo-mass distribution, allowing us to isolate clustering differences that are independent of halo mass. The samples yield approximately 8,000 groups above and below the 84th--16th percentiles and 16,600 groups above and below the 67th--33rd percentiles drawn from the full catalogue (50,497). The selections are mutually exclusive: a group in the bright sample cannot appear in the faint sample and vice versa. We used this selection to perform the analysis in Sects.~\ref{sec:3.1}--\ref{sec:3.5}.
\par
To go on to assess whether differences in baryon retention primarily drive the luminosity-dependent clustering signal (see Sect.~\ref{sec:3.3}), we repeated the analysis by splitting halos according to their gas content at fixed mass. Specifically, within the same narrow mass bins used for the $L_{X, 200}$ selection, we ranked halos by their gas fraction within $R_{200}$ and defined gas-rich and gas-poor subsamples using identical percentile cuts. We choose to select all gas particles within a sphere of radius $R_{200}$, without any further cuts in temperature or pressure, to trace the total matter assembly in halos explicitly. 

\textcolor{black}{ To test whether this signal is directly connected to halo assembly, we repeated the measurement after additionally matching halos simultaneously by mass and formation time. This allowed us to test whether the luminosity- and gas-fraction-dependent clustering signals are primarily driven by differences in halo assembly history, rather than by baryonic properties acting as independent secondary variables. We defined the halo formation time, $z_{50}$, as the redshift at which the main progenitor first reaches half of its final $z=0$ mass.}

\subsection{Clustering measurements and halo bias}
Dark matter halos, and the galaxies they host, do not trace the matter density field uniformly. Instead, they preferentially populate overdense regions, a tendency commonly referred to as bias. We can quantify this bias by defining any halo overdensity as
\begin{equation}
\delta_h(\mathbf{r},M) \equiv \frac{n(\mathbf{r},M)}{\bar{n}(M)} - 1 ,
\end{equation}
where $n(\mathbf{r},M)$ is the local number density of halos of mass, $M$, and $\bar{n}(M)$ is the  corresponding mean number density. The matter overdensity is defined analogously as $\delta_m(\mathbf{r}) = \rho(\mathbf{r})/\bar{\rho} - 1$.

On sufficiently large scales, where density fluctuations are small, variations in the halo density trace those of the matter field in an approximately proportional way,
\begin{equation}
\delta_h(\mathbf{r},M) \simeq b(M)\,\delta_m(\mathbf{r}) ,
\end{equation}
where $b(M)$ is the bias parameter. A higher value of $b$ indicates that halos of a given mass are more strongly clustered than the underlying matter distribution.

To quantify this bias in practice, we can measure clustering either through halo--halo correlations or by comparing halos directly to the underlying matter field. Rather than measuring halo-halo clustering, which is affected by the discrete nature of halo samples and associated shot noise, we instead characterised clustering using the halo--matter cross-spectrum. In Fourier space, the above relation implies
\begin{equation} \label{eq:Phm}
P_{hm}(k,M) \equiv b(M)\,P_{mm}(k),
\end{equation}
where $P_{mm}(k)$ is the matter power spectrum calculated at the same redshift. To efficiently calculate the power spectrum in Eq.~\ref{eq:Phm}, we used the \textsc{Pk\_library}\footnote{https://github.com/franciscovillaescusa/Pylians}. Firstly, we reconstruct the density fields and compute the power spectra on a $1024^3$ piecewise cubic spline mesh grid. Particles are assigned to the grid using a cloud-in-cell (CIC) mass-assignment scheme, obtaining a mass density field $\rho(\mathbf{x})$. For the group catalogue, we used the comoving positions of the full sample (complete above $M_{200,\min} = 10^{12.5}\,M_{\odot}$) and construct the grid for the bright and faint samples. The resulting number density fields can be converted into overdensity fields $\delta_{\mathrm{h}}$ by removing the contribution of the average density field in the cells. We repeated the same procedure for the full group sample to obtain a reference clustering measurement which we used to we compare the bright and faint subsamples against.
\par
We therefore estimated the mean halo bias from Eq.~\ref{eq:Phm} as the ratio $P_{hm}(k)/P_{mm}(k)$ on large linear scales, where this quantity becomes approximately scale-independent (i.e. $b_{\rm lin}$). Throughout this work, we chose to focus on a single simulation snapshot and omit the explicit redshift dependence, for clarity.
\par
We assessed the robustness of the large-scale bias measurements in two ways. First, we repeated the calculation of $b_{\rm lin}$ using different choices of the maximum linear wavenumber in the range $0.1 \leq k_{\rm lin}/(h\,{\rm cMpc}^{-1}) \leq 0.2$. The results are presented in Appendix~\ref{appendixA}. Second, we estimated the uncertainties using spatial jackknife resampling of the simulation volume. We divided the simulation volume into a $N_{jk} = 3\times3\times3$ grid (27 subvolumes) and recompute $b_{\rm lin}$, leaving out one subvolume at a time. Since the jackknife regions are not fully independent of the largest modes, the resulting significances should be interpreted as internal uncertainties within the finite simulation volume, rather than as a complete error budget that includes cosmic variance. For each pair of subsamples, we quantified the significance of the large-scale bias difference as
\begin{equation}
\label{eq:S_jk}
S_{\rm JK} =
\frac{\Delta b_{\rm lin}}{\sigma (\Delta b_{\rm lin})},
\end{equation}
where
\[
\Delta b_{\rm lin}(z) = b_{\rm lin|84th}(z) - b_{\rm lin| 16th}(z),
\]
is the difference between the large-scale bias of the upper and lower percentile samples. We calculate it by averaging $b$ over linear modes with $k \le k_{\rm lin}=0.2\,h\,\mathrm{cMpc}^{-1}$. In parallel, we define the variance $\sigma^2(\Delta b_{\rm lin})$ using jackknife resampling of the simulation volume, namely,
\[
\sigma^2(\Delta b_{\rm lin}) = \frac{N_{jk}-1}{N_{jk}} \sum_j\left(\Delta b_{\rm lin}^{(j)}(z)-\overline{\Delta b_{\rm lin}(z)}\right)^2, 
\]
where $j$ runs over all jackknife resamples, $\overline{\Delta b_{\rm lin}(z)}$ is the averaged single linear bias difference in resampling, and $N_{jk}$ is the number of jackknife subvolumes.

where
\begin{equation}
\Delta b_{\rm lin} =
b_{\rm lin}^{\rm bright} - b_{\rm lin}^{\rm faint}
\end{equation}
is the difference between the large-scale bias of the upper- and lower-percentile samples, while $\sigma (\Delta b_{\rm lin})$ is the jackknife uncertainty on this difference, calculated as the standard deviation.

\begin{figure*}
    \centering
    \subfloat[]{
        \includegraphics[width=0.48\linewidth]{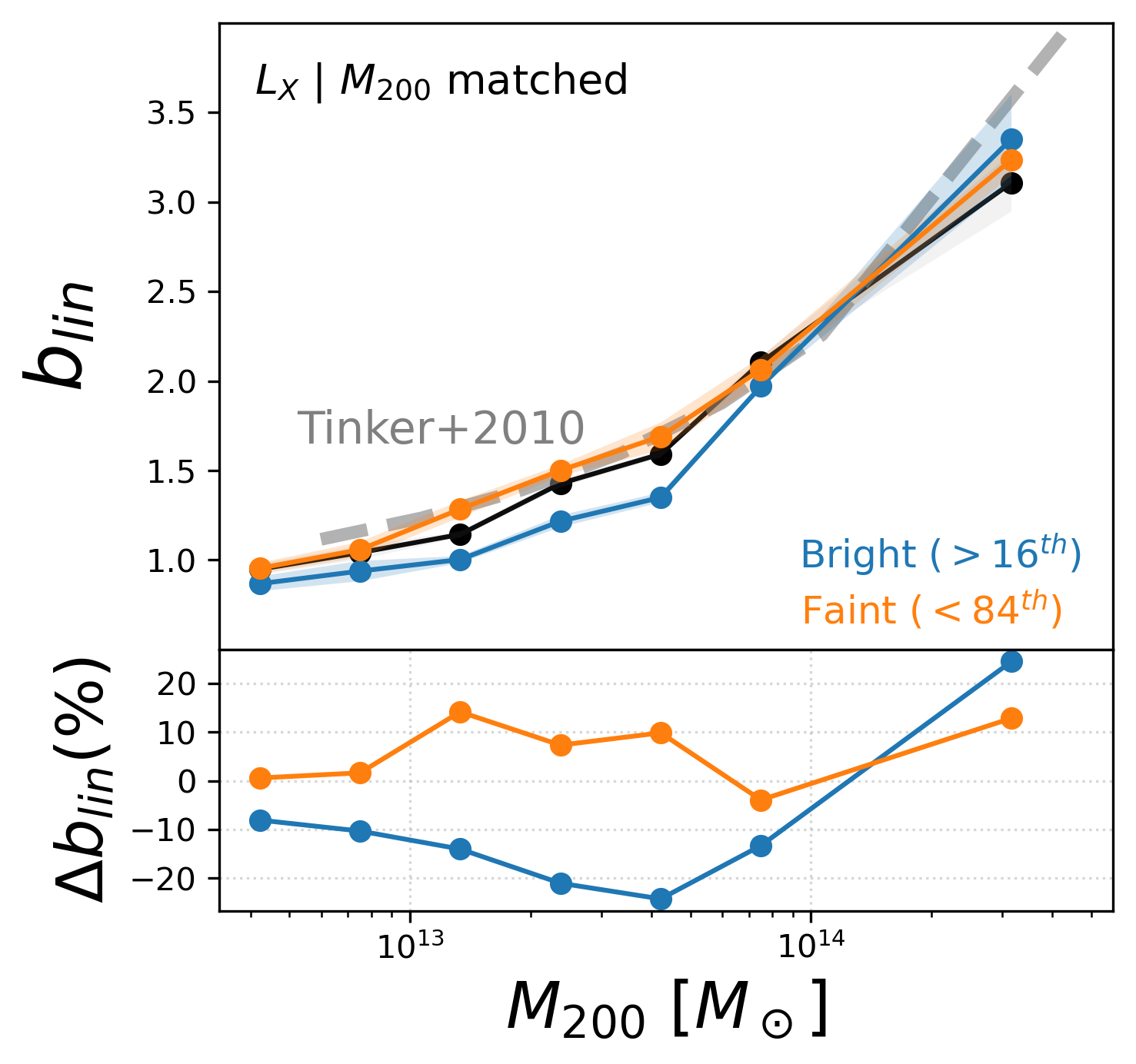}}
    \hfill
    \subfloat[]{
        \includegraphics[width=0.48\linewidth]{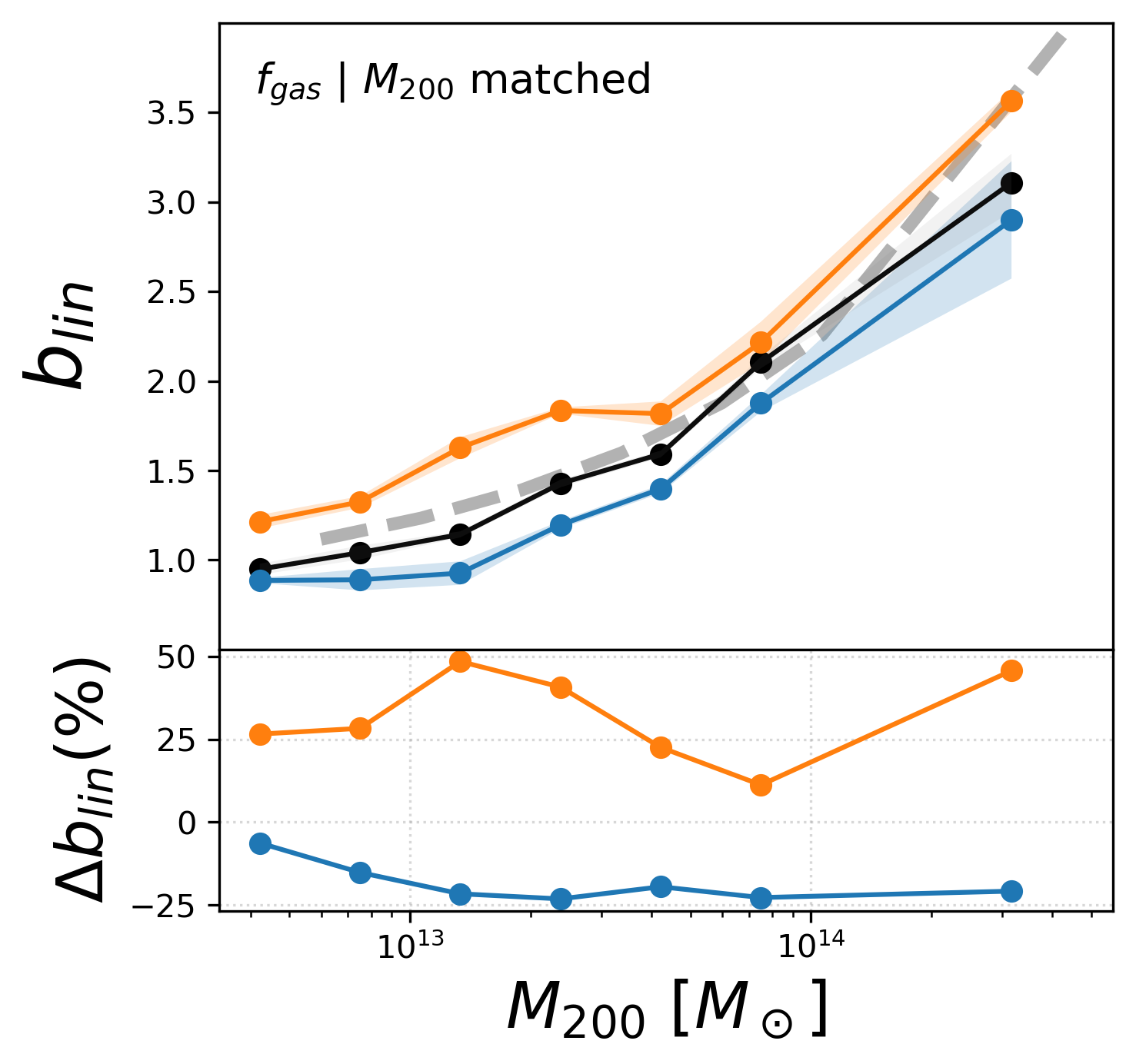}
    }
    \caption{Mean halo bias as a function of halo mass. Panel (a) Selection at fixed halo mass, with orange points indicating halos above the 84th percentile and blue points indicating halos below the 16th percentile. Panel (b) Corresponding selection in the total gas fraction. The black points represent the full catalogue. The grey dashed line is the baseline prediction from \protect\citet{tinker_large-scale_2010}. The shaded band represents the scatter in the relation when taking small differences in the value of the linear $k_{\rm lin}$, to account for the bias. Bottom panels: Percentage residual with respect to the full sample. }
    \label{fig:bias_M200}
\end{figure*}

\section{Results}
\label{sec:results}
\subsection{Clustering of halos selected on their X-ray luminosity}
\label{sec:3.1}
Figure~\ref{fig:Pk} illustrates the scale-dependent bias for the X-ray-bright and X-ray-faint halo samples following both the 84th-16th (solid lines) and 67th-33rd (dotted curves) percentiles split. Across the full range of wavenumbers probed, we find that X-ray-bright halos are more strongly clustered than X-ray-faint halos. This difference is already present on linear scales, where the bias is approximately scale independent, and becomes more pronounced (less scattered) toward smaller scales. 
\par
To quantify the total large-scale clustering difference, we extracted a single linear bias value, $b_{\rm lin}$, by averaging the scale-dependent bias, $b$, over the range $k < k_{\rm lin}$, where the bias is approximately constant. For the fiducial 84th-16th percentile split and $k_{\rm lin}=0.2\,h\,{\rm cMpc}^{-1}$, we find $ \Delta b_{\rm lin}=0.17$, corresponding to a $\sim17\%$ enhancement relative to the X-ray-faint sample. The jackknife uncertainty on this difference is $\sigma (\Delta b_{\rm lin})=0.03$, giving an internal jackknife significance of $S_{\rm JK}=5.12$. Similarly, the 67th-33rd percentile split gives $\Delta b_{\rm lin}=0.12$, corresponding to a $\sim12\%$ enhancement relative to the X-ray-faint sample, with $\sigma (\Delta b_{\rm lin})=0.02$ and $S_{\rm JK}=5.05$.

We checked the robustness of the value for $k_{\rm lin}$ by integrating for different evenly-spaced values of 0.01 between 0.1$\,h~\mathrm{cMpc}^{-1}$ and 0.2$\,h~\mathrm{cMpc}^{-1}$, resulting in negligible changes in the measured bias difference, although the formal jackknife significance increases mildly with increasing $k_{\rm lin}$ when more Fourier modes are included. Further details are given in Appendix~\ref{appendixA}. The shaded regions represent the uncertainty estimated via a spatial jackknife with a prefactor of $(N_{jk}-1)/N_{jk}$. Uncertainties are highest at the smallest wavenumbers, where the number of independent Fourier modes in the finite simulation volume is limited.

The persistence of the offset at low $k$ (large spatial scales) indicates a genuine difference in the large-scale bias of the two samples. Since the mass distributions are identical by construction, this result indicates that X-ray luminosity traces a secondary dependence of halo clustering beyond mass in the Magneticum simulation. As we discuss in Sect.~\ref{sec:4}, this secondary dependence can be understood as a baryonic imprint of halo assembly bias, which becomes visible in X-ray luminosity once the thermodynamic state of the gas is sufficiently coupled to halo growth history. On smaller scales, the enhanced bias of the bright sample reflects differences in internal structure and recent accretion activity, which affect the one-halo and transition regimes of the power spectrum.

\begin{figure}
    \centering    
    \subfloat{\includegraphics[width=\linewidth]{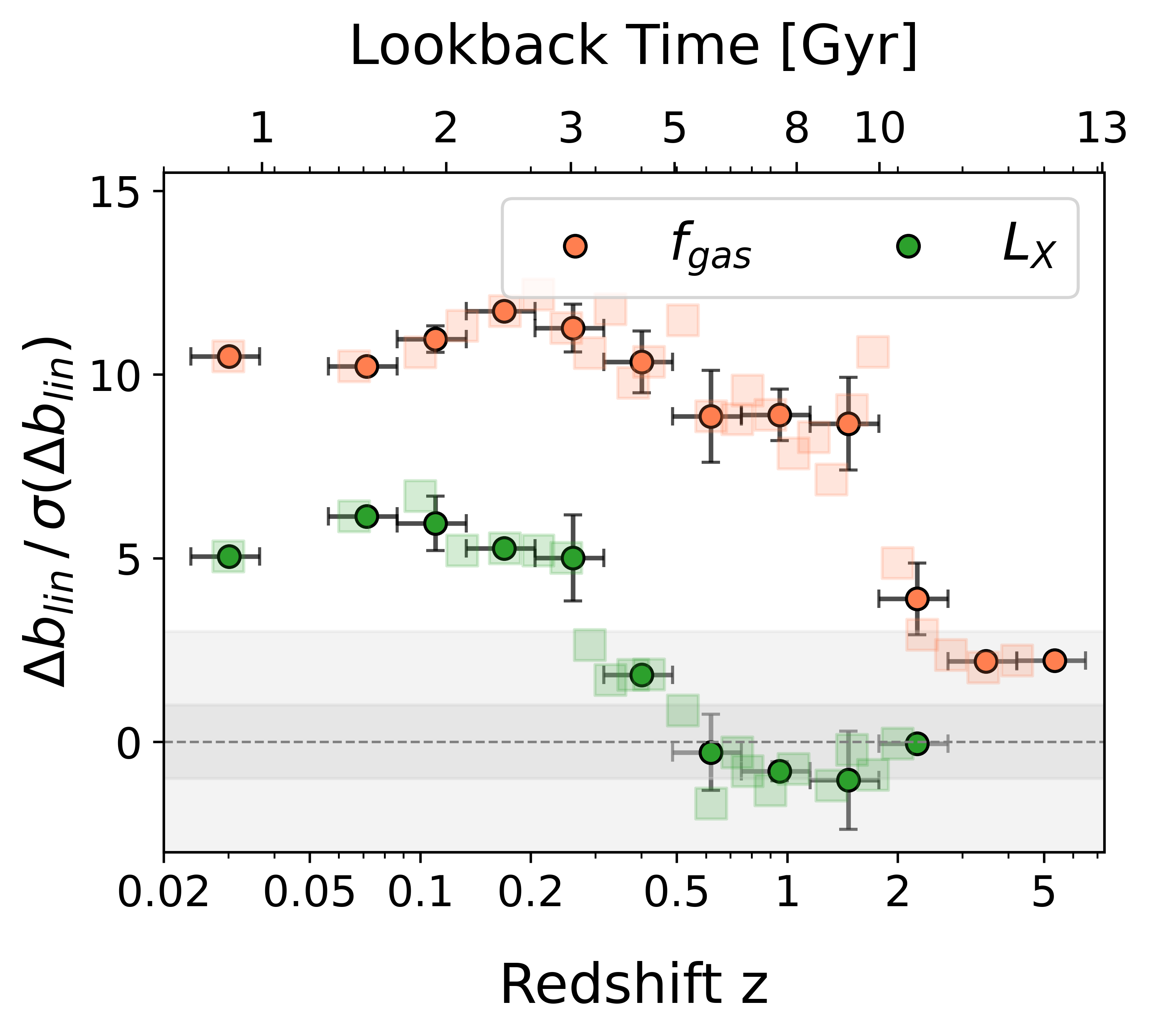}} 
    \caption{Significance of the large-scale clustering as a function of redshift. The round orange points show the difference in bias between the \textcolor{black}{ mass-matched} gas-rich and gas-poor halos across redshift, while the green round points correspond to the difference in the X-ray–bright and X-ray–faint systems. The underlying squared data points correspond to the measurements at each simulation snapshot, while the darker ones are the average over linearly spaced redshift bins. The error bars indicate the intrinsic scatter in the resampled bins, namely, the vertical range spanned by the significance in the given redshift bin and the bin size in the horizontal error bar. The gas-fraction–dependent signal persists across redshift, indicating an early imprint of baryonic assembly bias, whereas the X-ray–selected signal emerges only at low redshift, reflecting the late-time coupling between halo assembly and the thermodynamic state of the ICM. The shaded grey regions indicate the $\pm 1\sigma$ and $\pm 3\sigma$ signal-to-noise intervals. We neglect the clustering for the $L_X$ case beyond $z\sim2$, provided the very low signal-to-noise since $z\sim 0.5$. }
    \label{fig:significance_z}
\end{figure}

\subsection{Clustering dependence on mass}
\label{sec:3.2}
Furthermore, we calculated the linear bias as a function of halo mass and report the range spanned with the shaded bands in panel (a) of Fig.~\ref{fig:bias_M200}. At fixed mass, X-ray-bright halos consistently exhibit a higher linear bias than X-ray-faint halos. The relative difference in bias is of the order of $\sim$10--30\%, depending on the mass. It is interesting that the signal peaks at the galaxy group scales (i.e. $M_{200}\sim 10^{13}-10^{14}M_{\odot}$). In this regime, the scatter in gas content and assembly history is largest, AGN feedback is efficient at redistributing a substantial fraction of baryons from halo centres to radii comparable to and beyond $R_{200}$, and the associated back-reaction on the dark matter alters the halo density profile. Because groups are both abundant and feedback-efficient, they contribute strongly to the net suppression of $P_{mm}(k)$ on quasi- and non-linear scales. In our bias measurements, $b=P_{hm}(k)/P_{mm}(k)$, this implies that scale-dependent features at $k\gtrsim 0.3\,h\,\mathrm{cMpc}^{-1}$ should not be interpreted solely as changes in halo clustering, but also as the coupled response of the matter field to baryonic physics, which is expected to peak at group scales. The fact that our luminosity-dependent bias differences are largest in the group regime is therefore consistent with the expectation that this is where variations in gas retention and feedback-driven redistribution are maximal. We return to this point in Sect.~\ref{sec:4.1}, where we argue that the luminosity dependence emerges most clearly once halo gas has had sufficient time to thermalise and respond to the cumulative effects of accretion and feedback. Repeating the analysis using alternative luminosity cuts (67th--33rd percentiles) yields consistent results, demonstrating that the observed clustering trend is robust to the exact definition of the bright and faint subsamples.

As a baseline expectation for a purely mass-selected halo population, we compared our measurements to a standard mass-only halo bias calibration. In particular, we overplot the large-scale halo bias relation from \citet{tinker_large-scale_2010}, evaluated at the snapshot's redshift and for the same halo mass definition adopted in our catalogue. We also find good agreement with the bias model of~\citet{castro_impact_2021, Euclid:2024wog}, as expected given that it was calibrated on the same simulation suite using a similar methodology. We find that this baseline prediction aligns more closely with the X-ray-bright subsample, while the X-ray-faint halos exhibit a systematically lower bias. This behaviour indicates that the faint systems constitute a non-representative subset of the halo population, preferentially selecting early-forming, gas-poor halos whose large-scale clustering is suppressed relative to the mean. 

\textcolor{black}{ The agreement between the baseline model and the bright sample provides a useful consistency check on the clustering measurement. The offset of the faint sample indicates that in Magneticum, X-ray luminosity selects halos with a secondary dependence of halo bias beyond mass.} In the context of observations, this result also suggests that Malmquist-type selection effects in flux-limited X-ray surveys (where the most luminous systems at fixed mass are preferentially detected) might preferentially sample a population whose bias is closer to the mass-only expectation in Magneticum. However, the presence of a low-luminosity population with suppressed bias indicates that such selections still has the potential to introduce environmental trends that are not captured by mass-only halo bias models. \textcolor{black}{ Additionally, the connection between luminosity selection and clustering will be further modified by survey selection effects, flux limits, projection, aperture choices, and mass-observable scatter. Therefore, the comparison with the mass-only bias relation should not be interpreted as a direct prediction for flux-limited surveys. Instead, it would indicate that within Magneticum, the low-luminosity population carries most of the offset from the mass-selected expectation, while a realistic observational assessment requires forward modelling of the survey selection.}

\subsection{Clustering dependence on gas fraction}
\label{sec:3.3}
To assess whether differences in baryon retention primarily drive the luminosity-dependent clustering signal, we repeated the analysis by splitting halos according to their gas content at fixed mass, as discussed in Sect.~\ref{sec:2.3}. For the fiducial 84th--16th percentile split and $k_{\rm lin}=0.2\,h\,{\rm cMpc}^{-1}$, we found $\Delta b_{\rm lin}=0.39$, corresponding to a $\sim39\%$ enhancement relative to the gas-poor sample. The jackknife uncertainty on this difference is $\sigma (\Delta b_{\rm lin})=0.04$, giving an internal jackknife significance of $S_{\rm JK}=10.58$. Similarly, the 67th--33rd percentile split gives $\Delta b_{\rm lin}=0.27$, corresponding to a $\sim26\%$ enhancement relative to the gas-poor sample, with $\sigma (\Delta b_{\rm lin})=0.03$ and $S_{\rm JK}=10.53$. On large, linear scales, the separation in bias between gas-rich and gas-poor halos exceeds that observed for the X-ray-bright and faint samples. This is further clarified when investigating how the different mass scales impact the bias, as shown in panel (b) in Fig.~\ref{fig:bias_M200}. This suggests that gas content provides a more direct tracer of the assembly-dependent clustering signal than X-ray luminosity, which is additionally sensitive to the thermodynamic structure of the ICM. As discussed in Sect.~\ref{sec:4.1}, this difference is crucial: baryon retention preserves the memory of assembly earlier on, whereas X-ray luminosity becomes informative only once the gas has reached a more thermodynamically mature state.

At first glance, the stronger clustering of gas-rich halos is consistent with a simple picture in which halos in denser large-scale environments experience more sustained late-time accretion and thereby retain a larger fraction of their baryons. In this scenario, gas-poor systems correspond to halos that formed earlier or evolved in less dense regions, where gas accretion is less efficient. Internal processes such as AGN feedback, cooling, and stripping can, in principle, modulate the gas content and introduce additional scatter at fixed mass. However, our results indicate that, at low redshift, these processes do not erase the connection between baryon retention and environment. Instead, gas fraction remains strongly correlated with large-scale clustering, \textcolor{black}{ supporting the idea that the coupled evolution of dark matter and baryons can generate a measurable assembly-dependent clustering signal}. 

\subsection{Clustering dependence on redshift}
\label{sec:3.4}
We investigated the redshift evolution of the clustering signal by repeating the analysis at multiple simulation snapshots spanning the range $z\simeq 5$ to $z\simeq 0$. For each snapshot, we measured the halo--matter cross-power spectrum and derived the scale-dependent bias $b$ for the different subsamples. 

Figure~\ref{fig:significance_z} summarises the redshift dependence of the clustering signal in terms of the internal jackknife significance, $S_{\rm JK}$, defined in Eq.~\ref{eq:S_jk}. For the gas-fraction–selected samples, we find that gas-rich halos consistently exhibit a higher large-scale bias than gas-poor halos, from $z\simeq 2$. The sign of the signal is stable to lower redshift, and the bias difference is detected at the $> 10\sigma$ level. This indicates that the connection between baryon retention and large-scale environment is already established at early cosmic times and persists down to low redshift.

In contrast, the clustering signal obtained by splitting halos according to X-ray luminosity shows a more striking redshift dependence. At low redshift ($z\lesssim 0.3$), X-ray–bright halos are more strongly clustered than X-ray–faint systems, with a bias difference that is about half that of the gas-fraction split, but still detected at the $>5\sigma$ level. However, at $z\sim 0.5$, the luminosity-dependent signal rapidly weakens and becomes consistent with zero within the uncertainties. In this regime, $\Delta b_{\rm lin}$ does not exceed the $1\sigma$ threshold, and no statistically significant clustering difference is detected.

\begin{figure*}
    \centering    
    \subfloat{\includegraphics[width=\linewidth]{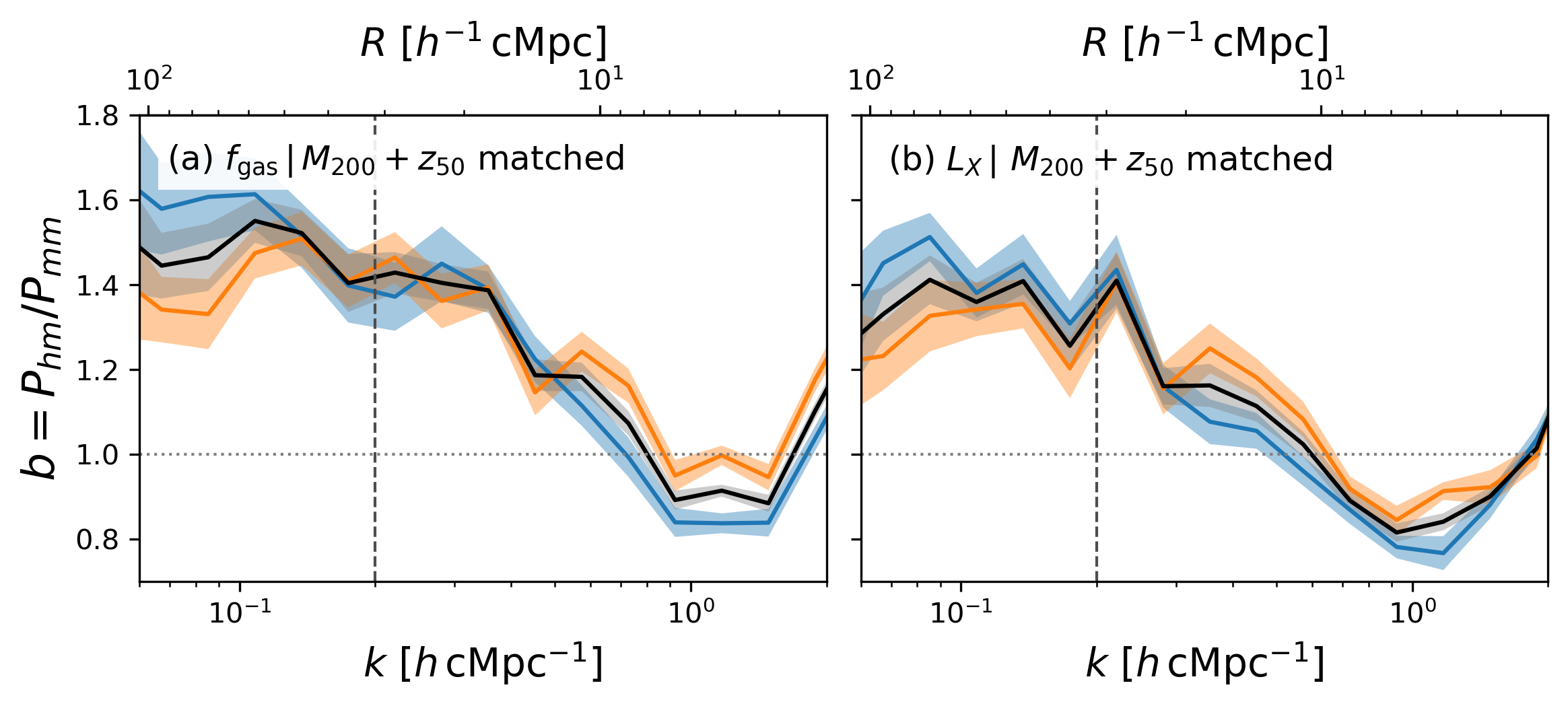}} 
    \caption{Scale-dependent bias for the full halo sample (black), faint (blue), and bright (orange) when mass and formation time matched. Panel (a) Selection based on the total gas fraction.  Panel (b) Selection based on the X-ray luminosity. Shaded regions indicate the $1\sigma$ scatter from jackknife resampling. The vertical dashed line marks the linear scale $k_{\rm lin}=0.2\,h\,{\rm cMpc}^{-1}$. The large-scale bias difference calculated in the two cases has significance below $2\sigma$.}
    \label{fig:Pk_z50}
\end{figure*}

\subsection{Formation-time matched samples}
\label{sec:3.5}

\textcolor{black}{ Having established the significance of the luminosity- and gas-fraction-dependent clustering signals, we then tested whether they are primarily driven by halo formation time. We repeated the selection after matching halos simultaneously in $M_{200}$ and $z_{50}$, where $z_{50}$ is defined as the redshift at which the main progenitor first reaches half of its final mass, as anticipated in Sect~\ref{sec:2.3}.}

\textcolor{black}{ After this additional matching, the large-scale bias difference is strongly reduced (see Fig.~\ref{fig:Pk_z50}). Although the best-fit scale-dependent bias shows an apparent change of sign at $k<0.2\,h\,{\rm cMpc}^{-1}$, the integrated large-scale difference has significance $S_{jk}<2\sigma$.  Thus, the formation time accounts for the original luminosity- and gas-fraction-dependent clustering signal in Magneticum. At $k>0.2\,h\,{\rm cMpc}^{-1}$, the original ordering of the samples is preserved, indicating that formation time mainly modifies the large-scale assembly-bias component, while smaller-scale clustering remains sensitive to additional halo and baryonic properties.}

\section{Discussion}
\label{sec:4}
\subsection{Interpreting the X-ray clustering}
\label{sec:4.1}
The clustering signal of X-ray-selected halos can be understood in terms of the evolving interplay between halo accretion, baryon retention, and the thermodynamic state of the ICM. The stronger gas-fraction signal, the group-scale peak, and the late emergence of the X-ray-luminosity dependence can all be understood as consequences of the evolving connection between halo accretion, baryon retention, and ICM thermodynamics. 

At high redshift, dark matter halos are in a fast-growth regime, characterised by rapid mass accretion and frequent mergers. In this phase, halo growth is dominated by recent infall, and the gaseous component is continuously replenished by newly accreted material that has not yet reached thermal equilibrium within the halo potential well. \textcolor{black}{ In Magneticum,} the total gas fraction remains closely connected to the halo assembly history and the surrounding large-scale environment. Halos embedded in denser regions continue to accrete gas efficiently and remain gas-rich, while halos in less dense environments are more susceptible to early gas depletion through feedback and stripping. This environmental dependence naturally gives rise to a difference in large-scale clustering between gas-rich and gas-poor systems, which is already detectable at high redshift.

In contrast, the thermodynamic state of the gas at early times is only weakly coupled to baryon retention. Temperature is strongly influenced by recent accretion shocks and merger-driven heating, rather than by the depth of the halo potential or by long-term gas confinement. As a result, gas-rich halos at high redshift typically host a substantial fraction of recently accreted, only partially thermalised gas and are therefore cooler on average than gas-poor halos, whose remaining gas has been more efficiently shock-heated and processed by early feedback. In this regime, gas fraction and temperature are not tightly correlated, and X-ray luminosity (given its sensitivity to both the gas density and temperature) does not provide a monotonic tracer of halo assembly or environment.

This explains why at high redshift, the clustering signal is clearly detected when halos are split by their total gas fraction, but it is weak or absent when using X-ray luminosity as a selection variable. Although a baryonic assembly bias is already present, it is not yet encoded in the X-ray properties of the ICM.

At a lower redshift, the situation changes. As the cosmic accretion rate declines, halos progressively exit the fast-growth regime and approach a more quasi-equilibrium state. Gas accretion slows down, merger activity becomes less frequent, and feedback processes (particularly AGN feedback) play an increasingly important role in regulating gas retention and redistribution \citep{castro_impact_2021}. In this phase, differences in assembly history and environment translate into systematic differences in both gas content and thermodynamic structure. Gas-rich halos tend to retain denser, cooler cores, while gas-poor halos host hotter and more diffuse gas distributions \citep[see Fig.~5--6 in][]{marini_impact_2025}.

Once halos reach this more thermodynamically mature stage, the X-ray luminosity becomes more tightly coupled to baryon retention and, indirectly, to halo assembly history. \textcolor{black}{ This interpretation is supported, but also complicated, by the formation time-matched test presented in Sect.~\ref{sec:3.5}. Once halos are compared at fixed mass and formation time, the integrated large-scale bias difference is reduced below $2\sigma$. The best-fit scale-dependent curves show an apparent sign change on the largest scales, but this is not statistically significant and should not be interpreted as evidence for a reversal of the effect. This suggests that X-ray luminosity might be a good proxy for halo formation time. Other correlated aspects of halo growth, such as recent accretion rate, concentration, merger history, tidal environment, or baryon-retention history, must also contribute to the residual signal.} As a result, the luminosity-dependent clustering signal emerges at low redshift, with X-ray--bright halos exhibiting a higher large-scale bias than X-ray--faint systems at fixed mass. The redshift evolution of the clustering signal therefore reflects a transition from an accretion-dominated regime, where baryon mass is the primary tracer of assembly bias, to a feedback-regulated regime, where X-ray observables encode the imprint of halo growth and environment.

\subsection{Observational limits}
\textcolor{black}{ A direct comparison with observations requires caution. The X-ray luminosities used here are intrinsic simulation-based quantities measured within $R_{200}$ for the gas component only. While the modelling includes spectral emission, redshift effects, and Galactic absorption (more details in Sect.~\ref{sec:2.3}), it does not include the full observational process that enters real X-ray catalogues. Observed luminosities are affected by projection effects, instrumental response, exposure variations, background subtraction, point-source contamination, aperture choices, flux limits, masking, and source-detection algorithms. In particular, extended-source detection depends not only on total flux, but also on surface brightness and extent likelihood \citep{brunner_erosita_2022}. As a result, low-surface-brightness or gas-poor systems might preferentially be missed \citep[e.g.][]{bulbul_erosita_2022}.}

\textcolor{black}{ A further limitation is that our selection was performed at a fixed true halo mass. In the observations, halo masses must be inferred from weak lensing, richness, dynamical measurements, SZ signal, or X-ray proxies, all of which have non-negligible scatter and might exhibit a covariance with the X-ray luminosity or gas fraction \citep{vikhlinin_chandra_2009, Okabe2010, saro_toward_2013, costanzi_cosmological_2021, grandis_srgerosita_2024}. Such a covariance can either dilute or enhance an apparent luminosity-dependent clustering signal. Similarly, the use of $R_{200}$ is idealised: in real data, this radius is inferred from an assumed mass proxy and cosmology, and luminosities are often measured within finite apertures and then extrapolated. These effects can couple aperture, mass calibration, and luminosity selection in a way that is absent from the intrinsic simulation analysis.}

\textcolor{black}{ Gas fractions are even harder to measure homogeneously for large group samples, since they depend on assumptions about gas profiles, hydrostatic equilibrium, clumping, extrapolation radius, temperature structure, and the separation of bound halo gas from projected foreground and background emission \citep{Ettori2009, eckert_low-scatter_2020, popesso_hot_2024}. Therefore, the luminosity- and gas-fraction-dependent clustering signals presented here should be interpreted as Magneticum-based physical predictions rather than direct predictions for observed catalogues. A robust observational comparison will require forward modelling through mock X-ray observations, including survey selection functions, measurement noise, mass-observable scatter, and the same source-detection and aperture definitions used in real X-ray surveys \citep[e.g.][]{seppi_modelling_2025, marini_detecting_2024, marini_detecting_2025, shreeram_quantifying_2025}.}

\section{Conclusions}
\label{sec:conclusions}

In this study, we investigate the large-scale clustering of galaxy groups and clusters as a function of their X-ray luminosity at fixed halo mass, using the Magneticum hydrodynamical simulation. Our analysis focuses on the halo–matter cross-power spectrum, which allows us to robustly quantify the halo bias, while minimising the impact of shot noise.

We constructed mass-matched subsamples of X-ray-bright and X-ray-faint halos by selecting systems in the upper and lower percentiles of the $L_{X,200}$–$M_{200}$ relation within narrow halo mass bins. This approach allowed us to isolate the dependence of clustering on X-ray luminosity independently of the halo mass. We measured the scale-dependent bias as $b=P_{hm}(k)/P_{mm}(k)$ and extracted a large-scale, approximately scale-independent bias by averaging over linear modes.

The physical origin of the luminosity-dependent clustering can be understood in the context of halo assembly bias. Previous works have shown that the scatter in X-ray luminosity at fixed mass reflects differences in halo assembly history, gas accretion, and merger activity, with Magneticum predicting X-ray-faint systems typically associated with earlier assembly and enhanced gas depletion \citep{marini_impact_2025}. The present analysis extends this picture by demonstrating that these assembly-related differences are also manifested in the large-scale clustering of halos at low redshift. In this sense, X-ray luminosity acts as an observable tracer of assembly-dependent halo bias.

Moreover, we show that the gas fraction in groups and clusters is a more informative tracer of assembly bias in Magneticum simulations. The difference in bias between similar-mass samples with distinct gas fractions is stable over the redshift range $z \sim 0-2$ and negligible at higher redshifts. This is in line with the findings of~\citet{Castro2024Euclid}, supporting the evidence that the gas fraction offers a high level of valuable information on the baryonic implications in groups and clusters.

\textcolor{black}{ We also tested whether the luminosity- and gas-fraction-dependent clustering signals are primarily driven by halo formation time $z_{50}$. After matching halos by both $M_{200}$ and $z_{50}$, the integrated large-scale bias difference between bright and faint is reduced to below $2\sigma$. Residual scale-dependent differences, at the non-linear scales (i.e. $k>0.2\,h\,{\rm cMpc}^{-1}$), suggest that other correlated halo and baryonic properties, such as recent accretion rates, concentration, merger history, tidal environment, gas distribution, or baryon-retention history, also offer a significant contribution.}

\textcolor{black}{ Taken together, these findings indicate a connection, within Magneticum, between the baryonic properties of halos and their large-scale clustering. The X-ray luminosity acts as an observable but indirect tracer of formation time at low redshift, while the gas fraction provides a more direct tracer of baryon retention. The effect might therefore be relevant for group-scale systems, which dominate the number counts of wide-area X-ray surveys such as eROSITA, but its observational impact must be quantified with realistic survey forward modelling before being incorporated into cosmological analyses of X-ray-selected samples. Accounting for the luminosity-dependent bias will therefore be important for forward-modelling approaches that jointly describe the mass--observable relation \citep{comparat_full-sky_2020, lau_x-raying_2025}, and the clustering properties of X-ray selected samples in cosmological analyses \citep[e.g.][]{pillepich_x-ray_2012, Seppi2024, clerc_srgerosita_2024,ghirardini_srgerosita_2024}.}

\begin{acknowledgements}
      The authors thank Alessandra Fumagalli for the scientific discussion on the cluster clustering and the anonymous referee for the discussion which improved the manuscript. This project has received funding from the European Research Council (ERC) under the European Union’s Horizon Europe research and innovation programme ERC CoG (Grant agreement No. 101045437) and by the Excellence Cluster ORIGINS2, which is funded by the Deutsche Forschungsgemeinschaft (DFG, German Research Foundation) under Germany’s Excellence Strategy- EXC-2094390783311. TC is supported by the Agenzia Spaziale Italiana (ASI) under - Euclid-FASE D Attivita' scientifica per la missione - Accordo attuativo ASI-INAF n. 2018-23-HH.0, by the National Recovery and Resilience Plan (NRRP), Mission 4, Component 2, Investment 1.1, Call for tender No. 1409 published on 14.9.2022 by the Italian Ministry of University and Research (MUR), funded by the European Union – NextGenerationEU– Project Title "Space-based cosmology with Euclid: the role of High-Performance Computing" – CUP J53D23019100001 - Grant Assignment Decree No. 962 adopted on 30/06/2023 by the Italian Ministry of University and Research (MUR); by the Italian Research Center on High-Performance Computing Big Data and Quantum Computing (ICSC), a project funded by European Union - NextGenerationEU - and National Recovery and Resilience Plan (NRRP) - Mission 4 Component 2, by the INFN INDARK PD51 grant, and by the PRIN 2022 project EMC2 - Euclid Mission Cluster Cosmology: unlock the full cosmological utility of the Euclid photometric cluster catalog (code no. J53D23001620006). KD acknowledges support by the COMPLEX project from the European Research Council (ERC) under the European Union’s Horizon 2020 research and innovation program grant agreement ERC-2019-AdG 882679. The calculations for the Magneticum simulations were carried out at the Leibniz Supercomputer Center (LRZ) under the project pr83li. 
\end{acknowledgements}

   \bibliographystyle{aa}
   \bibliography{references_downloaded}
\onecolumn
\begin{appendix}
\section{Dependence of linear bias on $k_{\rm lin}$}
\label{appendixA}

\begin{figure*}[hb!]
    \centering   \subfloat{\includegraphics[width=0.48\linewidth]{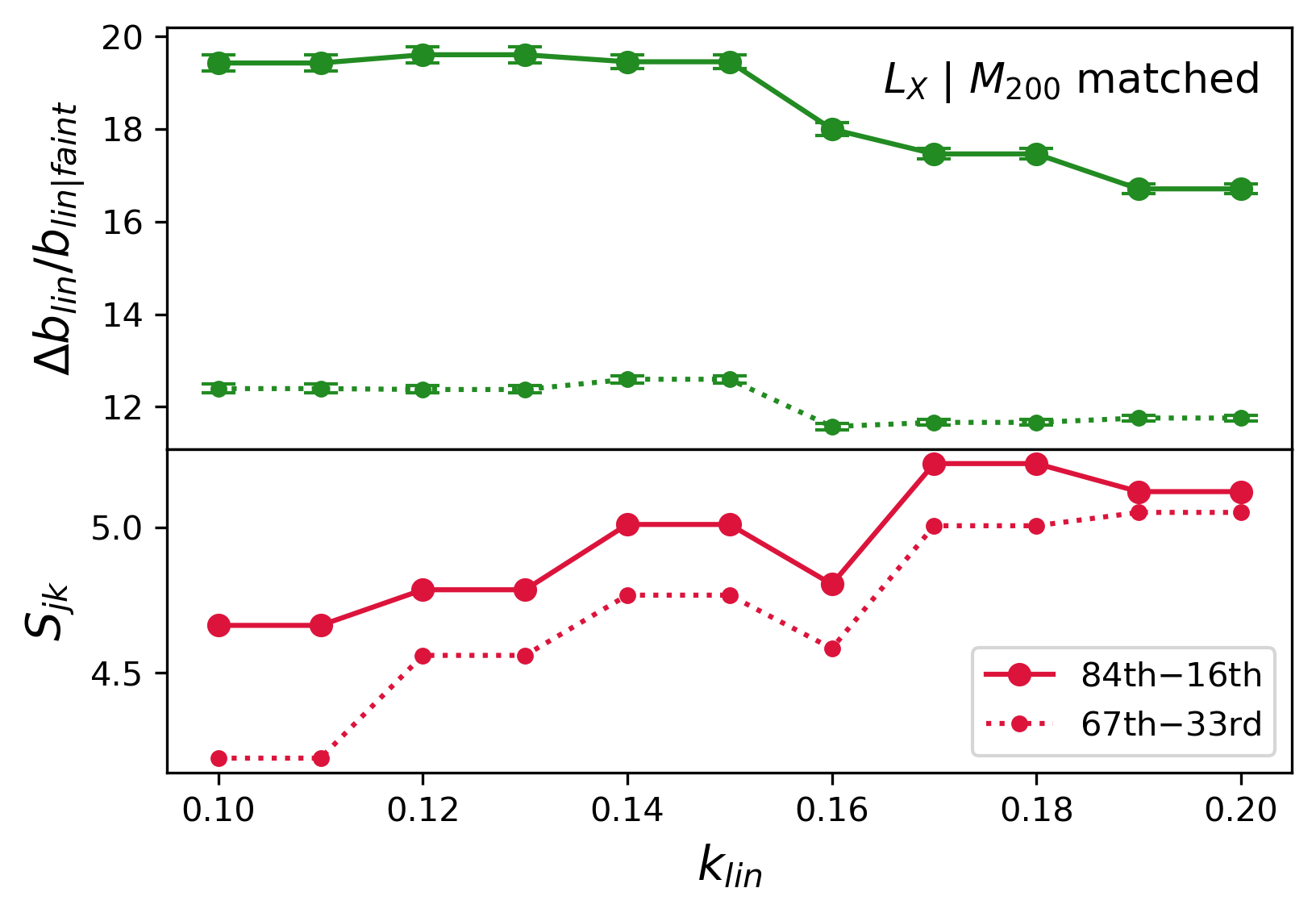}} 
    \subfloat{\includegraphics[width=0.48\linewidth]{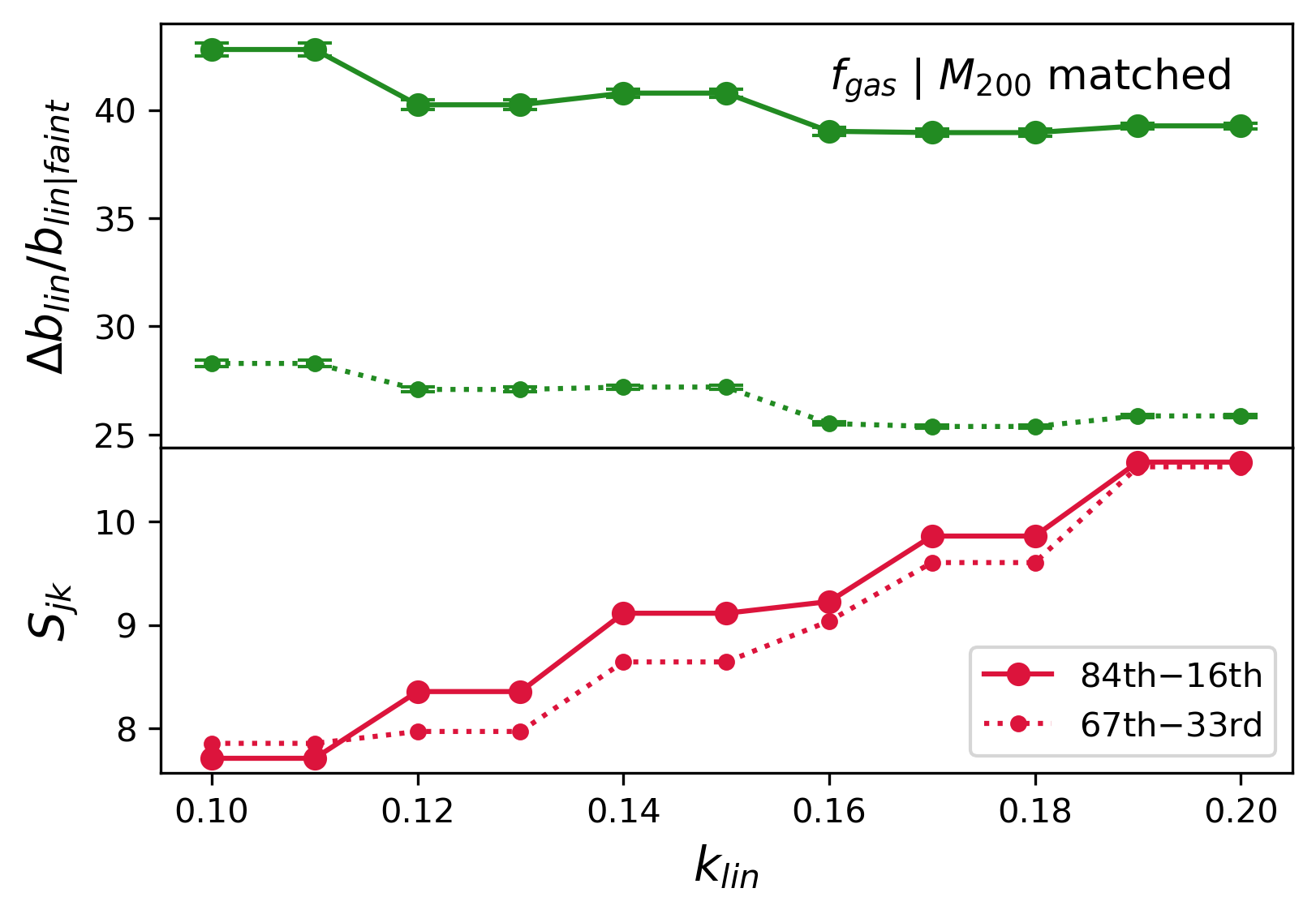}}
    \caption{Dependence of the X-ray luminosity (left) and gas-fraction (right) selected clustering signal on the adopted linear-scale cut. Top panel: Excess large-scale bias over the faint sample, $\Delta b_{\rm lin}/b_{lin|faint}$, expressed as a percentage, as a function of the maximum wavenumber $k_{\rm lin}$ used to define the linear bias. Bottom panel: Corresponding jackknife signal-to-noise ratio, $S_{\rm JK}=\Delta b_{\rm lin}/\sigma (\Delta b_{\rm lin})$, where $\sigma (\Delta b_{\rm lin})$ is estimated directly from the jackknife distribution. Solid and dotted curves correspond to the 84th--16th and 67th--33rd percentile splits, respectively. The positive bias difference and its high jackknife significance are stable over $0.1\leq k_{\rm lin}/(h\,{\rm cMpc}^{-1})\leq0.2$, showing that the gas-fraction-dependent clustering signal does not depend sensitively on the chosen linear-scale cut.}
    \label{fig:klin_dependence}
\end{figure*}

We tested whether the outcome of the linear bias difference between the faint and bright samples is impacted by the choice of $k_{\rm lin}$ when estimating the average. The steps in this procedure are as followed. First, we measure $\Delta b_{lin}$, which is  the difference in the linear bias between bright and faint samples, over the bias measured for the faint sample alone for values of $k_{\rm lin}$ between 0.1 and 0.2 with steps of 0.01. This provides a relative difference expressed in percentage with respect to the faint sample. Additionally, we estimate the robustness of the significance $S_{jk}$ from the jackknife. The results are presented in Fig.~\ref{fig:klin_dependence}. 
\par
Over the span of $k_{\rm lin}$ tested, the clustering signal does not vary significantly. The largest difference is among different percentile splits, which is expected from the clustering argument. In parallel, the significance measurement is increasing with increasing $k_{\rm lin}$, we find that at all times is $>3\sigma$.
\vspace{0.3cm}

\end{appendix}  

\end{document}